\documentclass[9pt,twocolumn,twoside]{pnas-new}

\templatetype{pnasresearcharticle} 

\usepackage{color}    
\usepackage{graphicx}
\usepackage{dcolumn}
\usepackage{bm}
\usepackage{subfigure}
\usepackage{amssymb}
\usepackage{multirow}
\usepackage{natbib}
    \usepackage{amsmath}
\graphicspath{{plots/}}
\usepackage{hyperref}
\hypersetup{
    colorlinks = true,
    citecolor = blue
}
\usepackage[utf8]{inputenc}
\usepackage[T1]{fontenc}
\usepackage[normalem]{ulem}

\newcommand{\sss}{\scriptscriptstyle}
\newcommand{\br}{\bm{r}}

\newcommand{\bq}{\bm{q}}

\newcommand{\s}{_\mathrm{{\scriptscriptstyle S}}}
\newcommand{\h}{_\mathrm{{\scriptscriptstyle H}}}
\newcommand{\xc}{_\mathrm{{\scriptscriptstyle XC}}}

\renewcommand{\vec}[1]{\mathbf{#1}}

\title{Towards a Quantum Fluid Theory of Correlated Many-Fermion Systems from First Principles}

\author[a,b]{Z. A. Moldabekov}
\author[a,b]{T. Dornheim} 
\author[c]{G. Gregori} 
\author[d]{F. Graziani} 
\author[e]{M. Bonitz} 
\author[a,b,1]{A. Cangi}

\affil[a]{Center for Advanced Systems Understanding (CASUS), D-02826 G\"orlitz, Germany}
\affil[b]{Helmholtz-Zentrum Dresden-Rossendorf, D-01328 Dresden, Germany}
\affil[c]{Department of Physics, University of Oxford, Parks Road, Oxford OX1 3PU, UK}
\affil[d]{Lawrence Livermore National Laboratory, Livermore, CA, 94550, USA}
\affil[e]{Institut f\"ur Theoretische Physik und Astrophysik, Christian-Albrechts-Universit\"at zu Kiel, Leibnizstra{\ss}e 15, 24098 Kiel, Germany}
\leadauthor{Moldabekov} 

\significancestatement{Many-fermion systems are ubiquitous in physics, and the inherent correlation among fermions greatly impacts macroscopic phenomena, for example, in high-energy-density physics and nanoplasmonics.
The quantum Bohm potential is the key quantity in quantum hydrodynamics and enables simulations of many-fermion systems at length and time scales that are not attainable with common first-principles methods.
In this work, we generate the very first exact Bohm potential of a many-fermion system — the many-fermion quantum Bohm potential. It will unlock unprecedented simulation capabilities in a multitude of applications where standard algorithms become infeasible. It might also impact cosmology.}

\authorcontributions{Z. A. Moldabekov performed the KS-DFT simulations, analyzed the results, contributed to derivations in the theory section and made a major contribution to writing the paper. T. Dornheim performed QMC simulations and contributed to writing the paper. G. Gregori contributed to writing the paper and provided insights on relevance of the results for cosmology. F. Graziani contributed to writing the paper and provided insights on the relevance of the results for hydrodynamics in high-energy-density plasmas. M. Bonitz contributed to derivation of the many-fermion Bohm potential and to writing the paper. A. Cangi made major contributions to writing the paper and contributed to derivations in the theory section.}
\authordeclaration{Authors have no competing interests.}
\correspondingauthor{\textsuperscript{1}To whom correspondence should be addressed. E-mail: a.cangi@hzdr.de}


\begin{abstract}
Correlated many-fermion systems emerge in a broad range of phenomena in warm dense matter, plasmonics, and ultracold atoms. Quantum hydrodynamics (QHD) complements common first-principles methods for many-fermion systems and enables simulations at larger length and longer time scales. 
While the quantum Bohm potential is central to QHD, we illustrate its failure for strong perturbations. We extend QHD to this regime by utilizing the \emph{many-fermion quantum Bohm potential} that is obtained from first-principles calculations. This opens up the path to more accurate simulations in strongly perturbed warm dense matter, inhomogeneous quantum plasmas, and on nano-structure surfaces at scales unattainable with first-principles algorithms. The \emph{many-fermion quantum Bohm potential} might also have important astrophysical applications in developing conformal-invariant cosmologies.
\end{abstract}

\dates{This manuscript was compiled on \today}
\doi{\url{www.pnas.org/cgi/doi/10.1073/pnas.XXXXXXXXXX}}

\begin{document}

\maketitle
\thispagestyle{firststyle}
\ifthenelse{\boolean{shortarticle}}{\ifthenelse{\boolean{singlecolumn}}{\abscontentformatted}{\abscontent}}{}

\dropcap{C}orrelated quantum many-fermion systems are currently in the focus of several fields ranging from high-energy-density physics~\cite{graziani-book} to ultracold fermionic atoms~\cite{Chien} and correlated materials~\cite{correlated-els}. Progress in all these fields relies on accurate theory and simulations including quantum Monte Carlo (QMC)~\cite{dornheim_physrep_18}, density functional theory (DFT)~\cite{hohenberg-kohn,runge-gross}, nonequilibrium Green functions~\cite{schluenzen_prb16}, and density matrix renormalization group (DMRG) methods~\cite{schluenzen_prb17}. While remarkable progress was achieved with these methods, their high computational cost and fundamental bottlenecks significantly restrict their application. For example, the fermion sign problem complicates the use of QMC~\cite{dornheim_sign_problem}, or the computational cost renders DMRG applications in three spatial dimensions infeasible. 
Therefore, there is a high need for complementary methods that extend the domain of simulations to length and time scales relevant for experiments, even at the price of reduced accuracy. 

One such method is quantum hydrodynamics (QHD). There has recently been a surge of activities in a number of research areas including warm dense matter (WDM)~\cite{Lardereaaw1634, bonitz_pop_19,PhysRevResearch.2.023036, RevModPhys.83.885}, plasmonics~\cite{10.1117/12.2320737,   PhysRevB.95.245434, toscano_nat-com_15}, electron transport in semiconductor devices and thin metal films~\cite{Manfredi_2001,  QDD1}, reactive scattering \cite{Wyatt, PRL_2001}, cosmology, and dark matter research~\cite{Gregori_2019, PERELMAN2019546,  PhysRevD.69.103512}.

QHD complements the aforementioned first-principles methods by enabling simulations at larger length and longer time scales. The quantum Bohm potential is central to QHD~\cite{Manfredi_2001, RevModPhys.83.885}. It captures quantum tunneling, spill out, and other non-local effects. Commonly, the quantum Bohm potential is approximated as
$v_B(\br, t) = 
-\hbar^2/(2m) \left[\nabla^2 \sqrt{n(\br, t)}/\sqrt{  n(\br, t)}\right]$ 
in terms of the mean density of electrons $n(\br, t)$; hereafter called \textit{standard Bohm potential}.
It is utilized in this form to model phenomena in various many-fermion systems. 

While standard QHD has proven useful, we question the validity of the standard Bohm potential when strong density perturbations are present. These emerge, for example, in strongly perturbed WDM~\cite{Dornheim_PRL_2020} and quantum plasmas \cite{Fletcher2015, PhysRevLett.112.105002}. Most notably, this regime is probed in recent and upcoming X-Ray scattering measurements of matter that is shock-compressed and laser-excited~\cite{Ofori_Okai_2018, Goulielmakis1267, fruehling_np_09, Kazansky_2019, Helled, mre16} using the seeding technique discussed in the conclusions.

In this research report, we therefore extend QHD to the regime of strong density perturbations. Our central result is to utilize the \emph{many-fermion quantum Bohm potential}
\begin{eqnarray}\label{eq:bohm-micro}
\tilde{v}_B(\br, t) &=& - \frac{\hbar^2}{2mN}\sum_{i=1}^{N} f_i \, \frac{\nabla^2 \sqrt{n_i(\br, t)}}{\sqrt{n_i(\br, t)}}
\end{eqnarray}
where $N$ is the total number of electrons. Specifically, we (1) generate an exact \emph{many-fermion quantum Bohm potential} based on \emph{exact} QMC data, (2) show how the standard Bohm potential breaks down for strong density perturbations, and (3) highlight how forces -- the key ingredient to QHD -- differ greatly in this regime. We highlight the practical importance of this result by turning our attention to a challenging many-fermion system -- the harmonically perturbed, interacting electron gas at finite temperature -- which is generated and probed in high-energy density physics facilities around the globe. 

Utilizing the \emph{many-fermion quantum Bohm potential} in QHD is motivated by the fact that it is derived from the exact quantum dynamics of electrons within time-dependent DFT~\cite{runge-gross} which provides the crucial link between QHD and interacting many-fermion systems.

\subsection*{Theory}
We begin with the non-relativistic, many-particle Hamiltonian of interacting fermions 
\begin{align}
\hat{H} = \hat{T} + \hat{V}_{\sss ee} + \hat{V},
\end{align}
where $\hat{T}$ denotes the kinetic energy operator, $\hat{V}_{\sss ee}$ the electron-electron interaction and $\hat{V}$ the external potential including the ionic background. The solutions are $N$-particle wave functions that are antisymmetric and normalized. For the sake of clarity we consider only spin-unpolarized systems.
A formally exact and computationally feasible solution to the quantum dynamics of electrons is given within time-dependent DFT~\cite{runge-gross}. Here, a set of $N$ time-dependent Kohn-Sham (KS) equations 
\begin{align} 
i\hbar \frac{\partial}{\partial t} \phi_i(\br, t) 
&= \left[-\frac{\hbar^2}{2m}\nabla^2 + v\s(\br, t)\right] \phi_i(\br, t)\,,
\label{eq:e_wave}
\end{align}
yields the exact time evolution of the electronic density, $n(\br, t) = \sum_i f_i |\phi_i(\br, t)|^2$, in terms of the single-particle KS orbitals $\phi_i(\br, t)$, where $f_i$ denotes an occupation function. This is achieved by the KS potential, $v\s(\br,t)= v(\br,t) + v\h[n](\br,t) + v\xc[n](\br,t)$, 
which exactly mimicks the electron-electron interaction within a mean-field description. 
Here, $v$ denotes the external potential, $v\h[n]$ the classical electrostatic (Hartree) potential, and $v\xc[n]$ the exchange-correlation potential.

Now, the time-dependent KS equations are reformulated into a set of coupled QHD equations by the following steps: 
(1) we insert the amplitude-phase representation of the KS orbitals~\cite{PhysRev.85.166}, 
$\phi_i(\br,t)=\sqrt{n_i(\br,t)}\exp\left[i{S_i(\br,t)}\right]$, 
into the time-dependent KS equations; 
(2) we use the expression for the mean orbital density, 
$\bar  n(\br,t) = \sum_i f_i n_i(\br,t)/N$, and velocity, $ {\vec v} = \sum_i f_i \vec v_i/N$, where we introduce $n_i(\br,t)=|\phi_i(\br, t)|^2$, as the KS orbital density and $\vec v_i=\vec \nabla S_i(\br,t)/m$, as the KS orbital velocity;
(3) we introduce density and velocity fluctuations $n_i=\bar n+\delta n_i$ and $\vec v_i={\vec v}+\delta \vec v_i$.
These steps yield the formally exact QHD equations
\begin{align}
 \frac{\partial  {\bar n}}{\partial t} &+\frac{1}{N}  \sum_i f_i \nabla\cdot (n_i\vec v_i) = 0\,,\label{eq:continuity}\\
 m \frac{\partial \vec v}{\partial t} &= - \nabla \tilde{v}_B 
-\frac{1}{n} \nabla P_e + \frac{1}{n}{\nabla \cdot {\bm \sigma}_e} + e \vec E-\vec \nabla v\xc\, \label{eq:momentum},
\end{align}
where we have not yet made any assumptions about velocity and density fluctuations.   
In Eq.~(\ref{eq:momentum}), $e$ is the absolute value of the electron charge, $P_e=\frac{1}{2m} \partial_\alpha\overline{\delta {p}_{i\alpha}^2}$ the electronic pressure term (with $\delta \vec p_i=m\delta \vec v_i$), ${\bm \sigma}_e=\frac{1}{m}\partial_\gamma \overline{\delta {p}_{i\alpha}\delta p_{i\gamma}}$ with $\gamma\ne\alpha$ the electronic viscous stress-tensor, and $\vec E=-\vec \nabla \left[ v + v\h \right]$ the electric field due to the Hartree and external potentials. The first equation is the continuity equation, whereas the second is the momentum conservation equation. Notice that the \emph{many-fermion quantum Bohm potential} emerges naturally. These QHD equations are equivalent to the time-dependent KS equations.

The QHD equations are turned into computationally feasible practice by employing approximations to (1) the exchange-correlation functional $v\xc$, 
(2) the equation of state $P_e$, (3) the viscous stress-tensor ${\bm \sigma}_e$, and (4) setting $\frac{1}{N}  \sum_i f_i \nabla\cdot (n_i\vec v_i) = \nabla \cdot (\bar n \vec v)$ in Eq.~(\ref{eq:continuity}) where the averaged fluctuations of a flux $\langle \delta \bar{\vec j}\rangle=\langle \delta n_i\delta \vec v_i\rangle$ are assumed to be negligible compared to the mean value $\bar{\vec{j}}=\bar n\vec v$. 
Using approximations to the equation of state and the viscous stress-tensor enables QHD to go beyond the length and time scales that are attainable in time-dependent DFT calculations. On the other hand, practical calculations spanning a large range of length and time scales are performed with classical hydrodynamics simulations. These, however, completely neglect quantum non-locality effects. As discussed below, these quantum effects become increasingly relevant for high-energy-density sciences due to ongoing and recent developments in experimental and diagnostic capabilities.

\begin{figure}\centering\includegraphics[width=0.48 \textwidth]{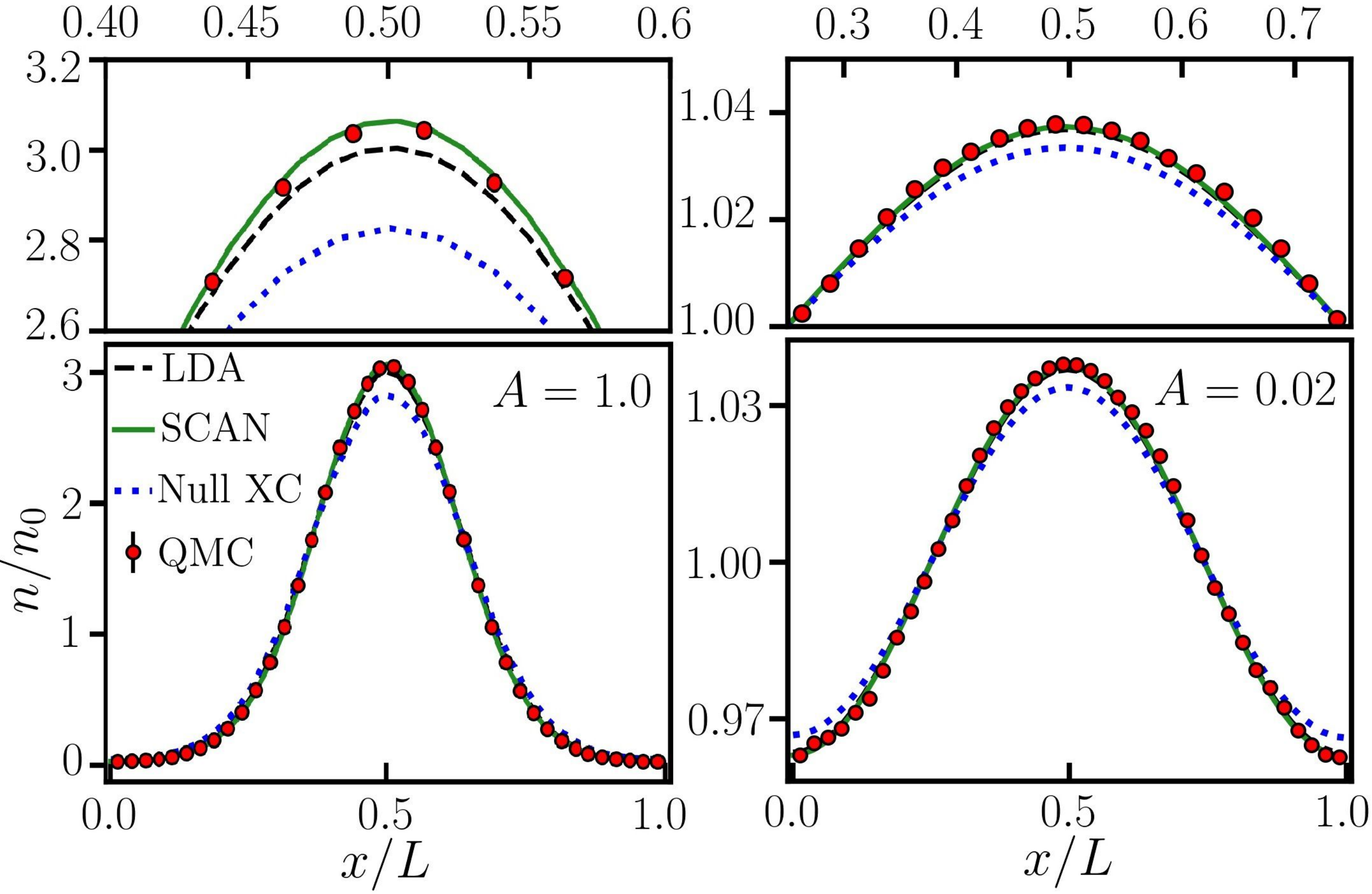}
\caption{\label{fig:nx}
Electronic density for two different amplitudes $A$, at $r_s=2$ and $\theta=1$. QMC results (red circles) are compared to KS-DFT data for different XC-potentials: solid green: SCAN; dashed black: LDA; dotted blue: non-interacting fermions ($v\xc=0$).}
\end{figure} 

\begin{figure*}
\center
\includegraphics[width=0.95 \textwidth]{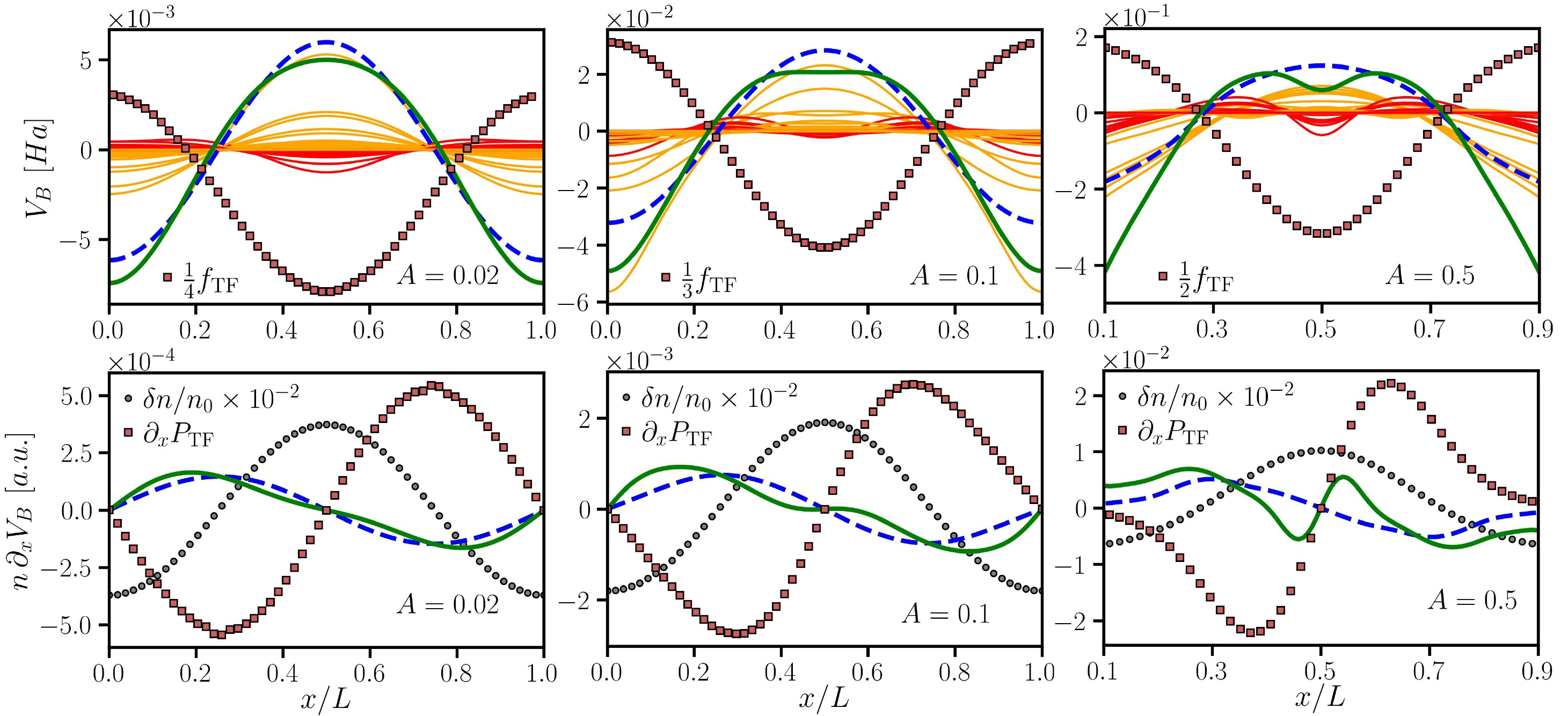}
\caption{\label{fig:Vx_A0_02}
Upper panel: Comparison of the exact \emph{many-fermion quantum Bohm potential} (thick green) with the standard Bohm potential (dashed blue) at $r_s=2$ and $\theta=1$. Additionally, the TF free energy density (red squares, scaled) and the contributing KS orbitals (thin red (dark) and orange (light) lines) are illustrated. The contribution of orbitals is scaled by a factor two (three) at $A=0.02$ ($A=0.1$ and $A=0.5$). Lower panel: Comparison of the forces from the \emph{many-fermion quantum Bohm potential} (green)  with forces from the standard Bohm potential (dashed blue). We also display the TF pressure (squares) and the density profile (grey circles). Note the scaling.}
\end{figure*} 

\subsection*{Results}
As the central result of this work we demonstrate the relevance of the \emph{many-fermion quantum Bohm potential} for the QHD equations (\ref{eq:continuity}) and (\ref{eq:momentum}), whereas in all prior works the standard Bohm potential was used.
First, we generate a \emph{many-fermion quantum Bohm potential} using KS-DFT based on exact QMC calculations of the harmonically perturbed, interacting electron gas. Then we show that the standard Bohm potential differs both qualitatively and quantitatively from $\tilde v_B$ to a great extent for strong density perturbations. 
Finally, we illustrate how these deviations yield vastly different forces. We, hence, argue that these lead to a different quantum plasma dynamics when used in the QHD equations. Agreement to better than $50\%$ in the resulting forces is achieved only for small density perturbations when $|\delta n|  \lesssim 10^{-3}\, n_0$ or $q>2\, q_F$. This is further analyzed in the Supplemental Material~\cite{supplement}.
The use of the \emph{many-fermion quantum Bohm potential} now renders QHD valid for the regime of strong density perturbations. While approximations to the pressure and viscous stress-tensor also influence the accuracy of the QHD equations, we focus on the \emph{many-fermion quantum Bohm potential}. It primarily determines the accurate inclusion of quantum effects, e.g., tunneling and spill-out, that are crucial for the aforementioned applications.

An important application that is highly relevant for high-energy-density physics is the harmonically perturbed, interacting electrons gas. It is described by the Hamiltonian
\begin{align}\label{eq:ham}
\hat{H} &= \hat{H}_{\rm UEG}+ \sum_{i=1}^N 2 A \cos\left(\br_i \cdot \bq \right)\,,
\end{align}
where $\hat{H}_{\rm UEG}$ denotes the Hamiltonian of the uniform electron gas with periodic boundary conditions. We choose the x-axis along $\mathbf{q}$ with $q=n q_{\rm min}$, $q_{\rm min}=2\pi/L$, $L=(N/n_0)^{1/3}$, and $n_0$ the number density of electrons.

The electronic states described by Eq.~(\ref{eq:ham}) are generated in recent WDM experiments (see Conclusions for further details). The amplitude $A$ in Eq.~(\ref{eq:ham}) controls the character of the KS orbitals. Tuning $A$ changes the the KS orbitals from plane waves to strongly localized wave packets. Moreover, by varying both $A$ and the wave number $q$, we tune the density gradients from small to large. The relevant parameter space is spanned by the density parameter $r_s=a/a_B$ and the degeneracy parameter $\theta=k_BT/E_F$, where $a$ is the mean inter-electronic distance, $a_B$ the first Bohr radius, $T$ the temperature, and $E_F$ the Fermi energy. For the remainder of this paper we choose $r_s=2$ and $\theta=1$. This corresponds to the WDM and quantum plasma regime \cite{graziani-book, dornheim_physrep_18}.

The construction of the \emph{many-fermion quantum Bohm potential} relies on accurate KS orbitals. We generate orbitals using KS-DFT for various amplitudes $10^{-3}\leq A\leq 1$ corresponding to the range from weak to strong perturbations. We assess their accuracy by comparing them with the exact result provided by QMC calculations.
The electronic densities for $A=1$ and $A=0.02$, using various exchange-correlation approximations (non-interacting fermions, LDA~\cite{KS1965:selfconsistent}, and SCAN~\cite{SCAN}) are illustrated in Fig.~\ref{fig:nx}, where $q_{\rm min}=0.84 q_F$. The comparison with the QMC data (red circles) confirms that the KS-DFT calculations using the SCAN functional provide the KS orbitals that virtually yield the exact density.

We now construct an exact \emph{many-fermion quantum Bohm potential} by inserting these KS orbitals into Eq.~(\ref{eq:bohm-micro}). The results are shown in the top panel of Fig.~\ref{fig:Vx_A0_02}. They are ordered in increasing perturbation strength ($A=0.02, 0.1, 0.5$). At the top we compare the \emph{many-fermion quantum Bohm potential} (thick green) with the standard Bohm potential (dashed blue). We observe significant differences for all amplitudes, and profound qualitative differences at high perturbation strength ($A=0.5$). To better understand the origin of these differences, consider contributions of the individual KS orbitals with a maximum and a minimum in the central region (orange and red lines). The former lead to a stronger \emph{many-fermion quantum Bohm potential} in the density depletion region at the edges, whereas the latter yield a weaker \emph{many-fermion quantum Bohm potential} in the central region where electrons accumulate. 
The important point to note is that the contribution from individual orbitals does not depend on the amplitude of the orbital density, but on its shape as is apparent from Eq.~(\ref{eq:bohm-micro}). This means that the contribution of a highly curved orbital can be critical, even if the corresponding occupation number may be relatively small.

Next, we relate these differences to the relevant energy scale in the QHD equations. We compare against the ideal part of the free energy density, $f_{\rm TF}[n(\vec r)]=\delta F_{\rm id}[n]/\delta n(\vec r)=\mu$ (red squares), which is a common approximation to the pressure in the QHD equations~\cite{ Manfredi_2001, RevModPhys.83.885} in terms of the Thomas-Fermi (TF) free-energy functional. The top panel of Fig.~\ref{fig:Vx_A0_02} shows that the ideal part of the free energy density has about the same order of magnitude as the quantum Bohm potential throughout highlighting the importance of the \emph{many-fermion quantum Bohm potential}. 

Now we assess the impact of using the \emph{many-fermion quantum Bohm potential}, instead of the standard Bohm potential, for simulating quantum dynamics.
We compute the force due to the pressure of a quantum Bohm potential from $n(\br) \nabla V_{B}$, where $V_B$ is the either the standard Bohm potential $v_B$ or the \emph{many-fermion quantum Bohm potential} $\tilde{v}_B$. To assess the importance of the observed differences, we compare them with the force due to TF pressure, $\vec  \nabla P_{\rm TF}=n(\br) \nabla f_{\rm TF}[n(\br)]$.
The lower panel of Fig.~\ref{fig:Vx_A0_02} demonstrates that the forces differ distinctly. At small perturbation strength the maximum deviation of the forces is already $50\%$. This deviation further increases with a stronger perturbation amplitude. At $A=0.5$, they differ substantially, and the standard Bohm potential fails to even yield a qualitative description. In the central region they are also qualitatively very different. Furthermore, the comparison with the TF force highlights the relative importance of the force due to $\tilde{v}_B$. At $A=0.02$, the TF force is about four times stronger than the force due to both variants of the quantum Bohm potential. With increasing perturbation strength, the force due to the \emph{many-fermion quantum Bohm potential} becomes more relevant. At $A=0.5$, it is close to the TF force in the central region, whereas it even exceeds the TF force in the density depletion regions close to the edges. 

\begin{figure}
\centering
\includegraphics[width=0.5 \textwidth]{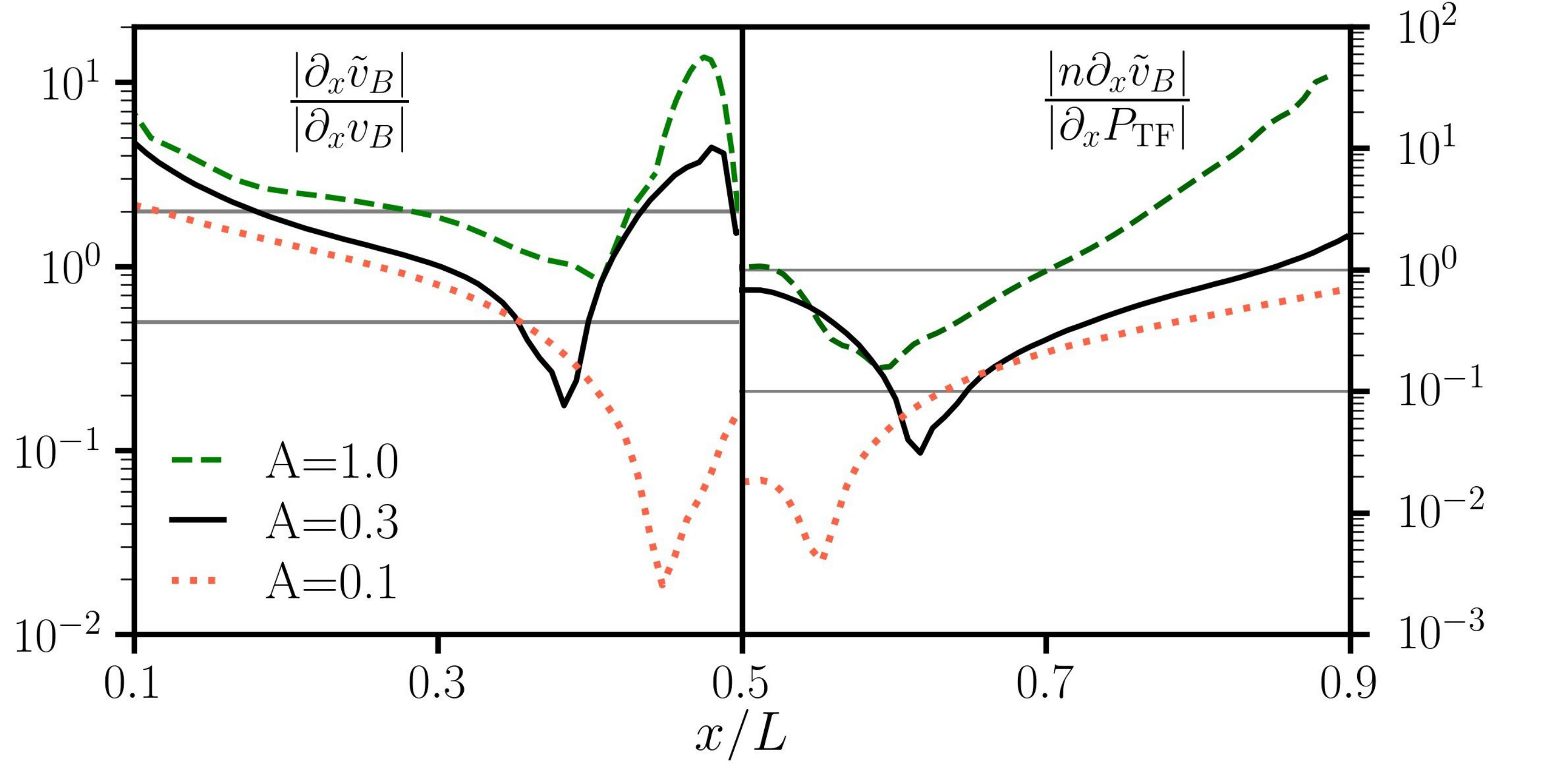}
\caption{\label{fig:VB-ratio}
Left: Ratio of the forces between the exact \emph{many-fermion quantum Bohm potential} ($\tilde{v}_B$) and the standard Bohm potential ($v_B$) at $r_s=2$ and $\theta=1$ for increasing density perturbation amplitudes $A$.  
Right: Ratio of the forces due to the quantum Bohm potential, $\tilde{v}_B$, and the TF pressure.}
\end{figure}

Next, in Fig.~\ref{fig:VB-ratio}, we provide a more detailed comparison of the forces. On the left, we show the ratio of the forces due to the \emph{many-fermion quantum Bohm potential} and the standard Bohm potential, whereas on the right, we show the ratio of the force due to $\tilde v_B$ with the TF force at $A=0.1,~0.3,~1.0$. We infer that, in general, the standard Bohm potential differs from the exact \emph{many-fermion quantum Bohm potential} by at least a factor of two throughout (left panel). For a small perturbation amplitude, $A=0.1$, the standard Bohm potential significantly overestimates (up to fifty times) the exact \emph{many-fermion quantum Bohm potential} in the central region and underestimates it by a factor of two in the density depletion region. At larger amplitudes ($A=0.3$ and $A=1.0$), the differences in both the density depletion region and in the central region increase. Finally, in Fig.~\ref{fig:VB-ratio} (right), we assess the relative importance of the quantum Bohm potentials. We deduce that the force due to the \emph{many-fermion quantum Bohm potential} is dominant in the density depletion regions, when $A\gtrsim 0.3$, with a maximum value of the density increase of $|\delta n|\gtrsim 0.6~n_0$. In conclusion, the \emph{many-fermion quantum Bohm potential} leads to a substantially different quantum dynamics of correlated many-fermion systems which has not been explored in any prior work. 

\subsection*{Conclusions and Outlook}
For a degenerate quantum many-particle system with Bose statistics in the condensate, the Madelung decomposition leads to the Gross-Pitaevski equation. Here, the exactness of the standard Bohm potential can be proven~\cite{bonitz_pop_19}. 
For fermionic systems, such a proof does not exist. A trivial exception is the case when the amplitudes of all orbitals coincide and the system is mapped onto a single orbital~\cite{bonitz_pop_19}. 

In this work, we carried out the very first investigation of the quantum Bohm potential for a correlated many-fermion system based on first-principles data from QMC and KS-DFT. Despite its long history in quantum mechanics since its derivation by Bohm in 1952~\cite{PhysRev.85.166} and its importance as a computational device in QHD, this has not been attempted before.
Our key result highlights the very limited applicability of the standard Bohm potential which is used in virtually all previous works of QHD. We showed that it is only valid for a very weakly perturbed electron gas (|$\delta n|  \lesssim 10^{-3}\, n_0$) or at very large wave-numbers ($q>2\, q_F$). Likewise, we demonstrated that the \emph{many-fermion quantum Bohm potential} is needed to model nonlinear phenomena in quantum plasmas and WDM. We further illustrated the significance of the force produced by the \emph{many-fermion quantum Bohm potential} for QHD simulations.

We anticipate that taking into account the \emph{many-fermion quantum Bohm potential} in quantum fluid approaches will play a significant role for many upcoming high-energy-density physics experiments. Strongly perturbed WDM states are generated and probed, for example, using ${\rm THz}$ lasers with an intensity of $600 ~{\rm kV/cm}$ that corresponds to a perturbation amplitude of $A\simeq 0.3$~\cite{Ofori_Okai_2018} and using free electron lasers with intensities of up to $10^{22}~{\rm W/cm^2}$ that lead to $A\approx 2$~\cite{Fletcher2015}. Likewise, it was recently demonstrated in an experiment~\cite{PhysRevX.8.031068} that spatially modulated WDM is created by laser pumping of a sample with a pre-designed, periodic grating structure. The induced WDM states can be characterized in-situ with the small-angle x-ray scattering technique using femtosecond X-Ray free-electron laser pulses on a spatial resolution of nanometers.  

Another exciting application of QHD is inertial confinement fusion~\cite{moses_national_2009} where strongly inhomogeneous electronic states emerge in the heating of shock-compressed fuel capsules. Of particular interest is the effect the \emph{many-fermion quantum Bohm potential} has on the shock behavior in high-energy density applications using lasers or pulsed power. The presence of higher-order spatial derivatives of the density produces a dissipative-like effect on the shock structure, shearing the interface and broadening the shock front. Other interesting applications include non-linear wave phenomena and instabilities in quantum plasmas~\cite{RevModPhys.83.885}. 
We also expect the \emph{many-fermion quantum Bohm potential} to impact the field of nano-plasmonics~\cite{10.1117/12.2320737,   PhysRevB.95.245434,toscano_nat-com_15} where simulations of large nano-clusters are routinely performed with QHD. Moreover, the \emph{many-fermion quantum Bohm potential} might enable quantum dynamics simulations of cold atom experiments that study transport properties~\cite{Chien}.
We also speculate that the force field generated by the \emph{many-fermion quantum Bohm potential} can be utilized as a computationally inexpensive neural-network surrogate model as it was done, e.g., for the free energy functional in KS-DFT~\cite{ellis2020accelerating, Brockherde2017BypassingTK} and the local field correction in QMC~\cite{Dornheim_PRL_2020_ESA}.

Finally, the \emph{many-fermion quantum Bohm potential} awaits exciting applications in cosmology. These approaches are based on an observation made by de Broglie pointing out that quantum mechanical effects are entirely equivalent to a conformal transformation of the background metric \cite{debroglie,Shojai}.
This leads to a representation of the non-local Bohm potential of all the particles in the Universe as an effective cosmological constant \cite{Gregori_2019}. Therefore, this outlines an interesting line of future research.

\matmethods{

\subsection*{KS-DFT simulation details} 
The KS-DFT calculations were performed  with GPAW~\cite{GPAW2}, which is a real-space implementation of the projector augmented-wave method.
A \emph{k}-point grid of $12\times 12\times 12$ using Monkhorst-Pack sampling of the Brillouin zone ($\vec k$-points) was used. At $\theta=1$, 180 orbitals (with the smallest occupation number of about $10^{-4}$) were used for a total of 14 electrons. The grid spacing was set to $0.15~{\rm \AA}$ for $10^{-3}\leq A\leq 1$ and $0.3~q_F \lesssim q\lesssim 2.53~q_F$.
The Hamiltonian of electrons is given by the sum of the standard (unperturbed) uniform electron gas Hamiltonian and the potential energy term corresponding to external perturbation.
Several exchange-correlation (XC) functionals were used: the standard LDA functional by Perdew-Zunger for degenerate electrons~\cite{Perdew_LDA}, the GDSMFB functional which is a parametrization of the LDA of the homogeneous electron gas at finite temperature~\cite{dornheim_prl16}, PBE~\cite{PBE}, PBEsol~\cite{PBEsol}, AM05~\cite{AM05}, and the meta-GGA functional SCAN~\cite{SCAN}.
The ab-initio quality of the KS-DFT calculations was validated with first-principles quantum Monte Carlo (QMC) calculations. We found that SCAN reproduces the exact QMC data more accurately than any of the other XC functionals. Only in the limit of a weak perturbation, the tested XC functionals yield agreement, as they reduce to the zero-temperature limit of the LDA. 

The relative error of the results obtained using different XC functionals compared to the QMC data is given in Fig.~\ref{fig:dn} for $A=1$ and $A=0.02$. The corresponding total density is presented Fig.~1 of the manuscript. The case $A=1$ corresponds to a strong-perturbation regime with a minimum density of $n\simeq 0.03 ~n_0$ close to the edges of the simulation box and with a maximum density of $n\simeq 3 ~n_0$ in the center. The case $A=0.02$ corresponds to the weak-perturbation regime with a density maximum $n\simeq 1.04 ~n_0$ and minimum $n\simeq 0.96 ~n_0$. From Fig.~\ref{fig:dn} we infer that the inclusion of XC effects by using the LDA significantly improves upon the fully non-interacting case (no XC functional) with a maximum error of about $4.7\%$ in the central region for $A=1$. However, using GGA functionals does not improve over the LDA results. The exact QMC data are reproduced remarkably well by the SCAN functional with an accuracy better than $1.43~\%$. We conclude that it is crucial to go beyond both LDA and GGA in order to obtain an accurate density when the perturbation is strong. A further analysis of this observation will be presented elsewhere.\\ 
 
\begin{figure}\vspace{0.5cm}\centering\includegraphics[width=0.46 \textwidth]{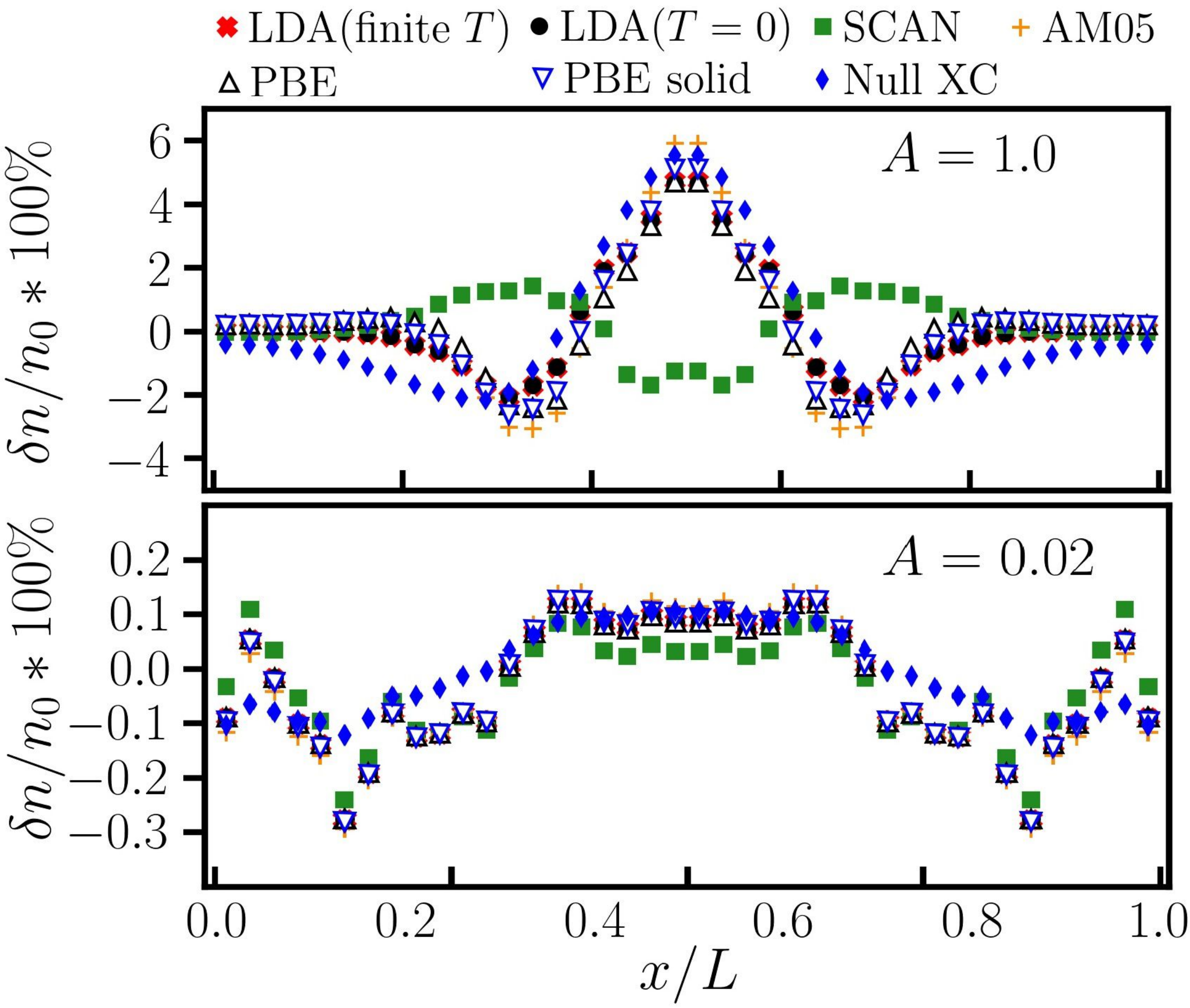}
\caption{
Relative error of the density using different XC functionals compared to the QMC data at $\theta=1$ and $r_s=2$.}
\label{fig:dn}
\end{figure}

\subsection*{QMC simulation details}
We use the standard path integral Monte Carlo (PIMC) method~\cite{dornheim_physrep_18} without any nodal restrictions on the thermal density matrix. Therefore, the simulations are computationally expensive due to the fermion sign problem~\cite{dornheim_sign_problem}, but exact within the given Monte Carlo error bars. We have used $P\sim10^2$ primitive imaginary-time propagators, which is fully sufficient to ensure convergence at  these parameters.
Additional details on the simulation of the harmonically perturbed electron gas at finite temperature can be found in Refs.~\cite{dornheim_physrep_18}.
}

\showmatmethods{} 

\acknow{This work was partly funded by the Center for Advanced Systems Understanding (CASUS) which is financed by Germany's Federal Ministry of Education and Research (BMBF) and by the Saxon Ministry for Science, Culture and Tourism (SMWK) with tax funds on the basis of the budget approved by the Saxon State Parliament.
We gratefully acknowledge CPU-time at the Norddeutscher Verbund f\"ur Hoch- und H\"ochstleistungsrechnen (HLRN) under grant shp00026 and on a Bull Cluster at the Center for Information Services and High Performace Computing (ZIH) at Technische Universit\"at Dresden.
The work of GG was funded in parts by the Engineering and Physical Sciences Research Council (grant numbers EP/M022331/1 and EP/N014472/1) and AWE plc. MB acknowledges support from the Deutsche Forschungsgemeinschaft via grant BO1366/15. A part of this work was performed under the auspices of the US Department of Energy by Lawrence Livermore National Laboratory under Contract DE-AC5207NA27344.}

\showacknow{} 

\bibliography{ref,mb-ref}

\begin{thebibliography}{10}

\bibitem{graziani-book}
F Graziani, MP Desjarlais, R Redmer, SB Trickey, {\em Frontiers and Challenges
  in Warm Dense Matter}.
\newblock (Springer), (2014).

\bibitem{Chien}
C Chien, S Peotta, M Di~Ventra, Quantum transport in ultracold atoms.
\newblock {\em\protect\JournalTitle{Nature Phys.}} \textbf{11}, 998–1004
  (2015).

\bibitem{correlated-els}
C Giannetti, et~al., Ultrafast optical spectroscopy of strongly correlated
  materials and high-temperature superconductors: a non-equilibrium approach.
\newblock {\em\protect\JournalTitle{Advances in Physics}} \textbf{65}, 58--238
  (2016).

\bibitem{dornheim_physrep_18}
T Dornheim, S Groth, M Bonitz, The uniform electron gas at warm dense matter
  conditions.
\newblock {\em\protect\JournalTitle{Phys. Rep.}} \textbf{744}, 1 -- 86 (2018).

\bibitem{hohenberg-kohn}
P Hohenberg, W Kohn, Inhomogeneous electron gas.
\newblock {\em\protect\JournalTitle{Phys. Rev.}} \textbf{136}, B864--B871
  (1964).

\bibitem{runge-gross}
E Runge, EKU Gross, Density-functional theory for time-dependent systems.
\newblock {\em\protect\JournalTitle{Phys. Rev. Lett.}} \textbf{52}, 997--1000
  (1984).

\bibitem{schluenzen_prb16}
N Schl{\"u}nzen, S Hermanns, M Bonitz, C Verdozzi, Dynamics of strongly
  correlated fermions:\textit{Ab initio} results for two and three dimensions.
\newblock {\em\protect\JournalTitle{Phys. Rev. B}} \textbf{93}, 035107 (2016).

\bibitem{schluenzen_prb17}
N Schl\"unzen, JP Joost, F Heidrich-Meisner, M Bonitz, {Nonequilibrium dynamics
  in the one-dimensional Fermi-Hubbard model: Comparison of the nonequilibrium
  Green-functions approach and the density matrix renormalization group
  method}.
\newblock {\em\protect\JournalTitle{Phys. Rev. B}} \textbf{95}, 165139 (2017).

\bibitem{dornheim_sign_problem}
T Dornheim, Fermion sign problem in path integral {M}onte {C}arlo simulations:
  Quantum dots, ultracold atoms, and warm dense matter.
\newblock {\em\protect\JournalTitle{Phys. Rev. E}} \textbf{100}, 023307 (2019).

\bibitem{Lardereaaw1634}
B Larder, et~al., Fast nonadiabatic dynamics of many-body quantum systems.
\newblock {\em\protect\JournalTitle{Science Advances}} \textbf{5} (2019).

\bibitem{bonitz_pop_19}
M Bonitz, ZA Moldabekov, TS Ramazanov, Quantum hydrodynamics for plasmas—quo
  vadis?
\newblock {\em\protect\JournalTitle{Physics of Plasmas}} \textbf{26}, 090601
  (2019).

\bibitem{PhysRevResearch.2.023036}
J Dufty, K Luo, J Wrighton, Generalized hydrodynamics revisited.
\newblock {\em\protect\JournalTitle{Phys. Rev. Research}} \textbf{2}, 023036
  (2020).

\bibitem{RevModPhys.83.885}
PK Shukla, B Eliasson, Colloquium: Nonlinear collective interactions in quantum
  plasmas with degenerate electron fluids.
\newblock {\em\protect\JournalTitle{Rev. Mod. Phys.}} \textbf{83}, 885--906
  (2011).

\bibitem{10.1117/12.2320737}
G Manfredi, PA Hervieux, F Tanjia, {Quantum hydrodynamics for nanoplasmonics}
  in {\em Plasmonics: Design, Materials, Fabrication, Characterization, and
  Applications XVI}, eds.{} DP Tsai, T Tanaka.
\newblock (International Society for Optics and Photonics, SPIE), Vol.{} 10722,
  pp. 12 -- 16 (2018).

\bibitem{PhysRevB.95.245434}
C Cirac\`{\i}, Current-dependent potential for nonlocal absorption in quantum
  hydrodynamic theory.
\newblock {\em\protect\JournalTitle{Phys. Rev. B}} \textbf{95}, 245434 (2017).

\bibitem{toscano_nat-com_15}
G Toscano, et~al., Resonance shifts and spill-out effects in self-consistent
  hydrodynamic nanoplasmonics.
\newblock {\em\protect\JournalTitle{Nat. Communications}} \textbf{6}, 7132
  (2015).

\bibitem{Manfredi_2001}
G Manfredi, F Haas, Self-consistent fluid model for a quantum electron gas.
\newblock {\em\protect\JournalTitle{Phys. Rev. B}} \textbf{64}, 075316 (2001).

\bibitem{QDD1}
D P., Mehats, F., R C., Quantum energy-transport and drift-diffusion models.
\newblock {\em\protect\JournalTitle{J Stat Phys}} \textbf{118}, 625--667
  (2020).

\bibitem{Wyatt}
RE Wyatt, {\em Quantum Dynamics with Trajectories Introduction to Quantum
  Hydrodynamics}, Interdisciplinary Applied Mathematics.
\newblock (Springer-Verlag New York) Vol.{}~28, (2005).

\bibitem{PRL_2001}
A Donoso, CC Martens, Quantum tunneling using entangled classical trajectories.
\newblock {\em\protect\JournalTitle{Phys. Rev. Lett.}} \textbf{87}, 223202
  (2001).

\bibitem{Gregori_2019}
G Gregori, B Reville, B Larder, Modified friedmann equations via conformal
  bohm{\textendash}de broglie gravity.
\newblock {\em\protect\JournalTitle{The Astrophysical Journal}} \textbf{886},
  50 (2019).

\bibitem{PERELMAN2019546}
CC Perelman, Bohm's potential, classical/quantum duality and repulsive gravity.
\newblock {\em\protect\JournalTitle{Physics Letters B}} \textbf{788}, 546 --
  551 (2019).

\bibitem{PhysRevD.69.103512}
PF Gonz\'alez-D\'{\i}az, Subquantum dark energy.
\newblock {\em\protect\JournalTitle{Phys. Rev. D}} \textbf{69}, 103512 (2004).

\bibitem{Dornheim_PRL_2020}
T Dornheim, J Vorberger, M Bonitz, Nonlinear electronic density response in
  warm dense matter.
\newblock {\em\protect\JournalTitle{Phys. Rev. Lett.}} \textbf{125}, 085001
  (2020).

\bibitem{Fletcher2015}
LB Fletcher, et~al., Ultrabright x-ray laser scattering for dynamic warm dense
  matter physics.
\newblock {\em\protect\JournalTitle{Nature Photonics}} \textbf{9}, 274--279
  (2015).

\bibitem{PhysRevLett.112.105002}
U Zastrau, et~al., Resolving ultrafast heating of dense cryogenic hydrogen.
\newblock {\em\protect\JournalTitle{Phys. Rev. Lett.}} \textbf{112}, 105002
  (2014).

\bibitem{Ofori_Okai_2018}
BK Ofori-Okai, et~al., A terahertz pump mega-electron-volt ultrafast electron
  diffraction probe apparatus at the {SLAC} accelerator structure test area
  facility.
\newblock {\em\protect\JournalTitle{J. Inst}} \textbf{13}, P06014--P06014
  (2018).

\bibitem{Goulielmakis1267}
E Goulielmakis, et~al., Direct measurement of light waves.
\newblock {\em\protect\JournalTitle{Science}} \textbf{305}, 1267--1269 (2004).

\bibitem{fruehling_np_09}
U Fr{\"u}hling, et~al., Single-shot terahertz-field-driven x-ray streak camera.
\newblock {\em\protect\JournalTitle{Nature Photonics}} \textbf{3}, 523 (2009).

\bibitem{Kazansky_2019}
AK Kazansky, IP Sazhina, NM Kabachnik, Angular streaking of auger-electrons by
  {THz} field.
\newblock {\em\protect\JournalTitle{J. Phys. B: Atomic, Mol. and Opt. Phys}}
  \textbf{52}, 045601 (2019).

\bibitem{Helled}
R Helled, G Mazzola, R Redmer, Understanding dense hydrogen at planetary
  conditions.
\newblock {\em\protect\JournalTitle{Nat. Rev. Phys.}} \textbf{2}, 562--574
  (2020).

\bibitem{mre16}
BY Sharkov, DH Hoffmann, AA Golubev, Y Zhao, High energy density physics with
  intense ion beams.
\newblock {\em\protect\JournalTitle{Matter and Radiation at Extremes}}
  \textbf{1}, 28--47 (2016).

\bibitem{PhysRev.85.166}
D Bohm, A suggested interpretation of the quantum theory in terms of "hidden"
  variables. i.
\newblock {\em\protect\JournalTitle{Phys. Rev.}} \textbf{85}, 166--179 (1952).

\bibitem{supplement}
Supplemental material, Technical report (year?).

\bibitem{KS1965:selfconsistent}
W Kohn, LJ Sham, Self-{{Consistent Equations Including Exchange}} and
  {{Correlation Effects}}.
\newblock {\em\protect\JournalTitle{Phys. Rev.}} \textbf{140}, A1133--A1138
  (1965).

\bibitem{SCAN}
J Sun, A Ruzsinszky, JP Perdew, Strongly constrained and appropriately normed
  semilocal density functional.
\newblock {\em\protect\JournalTitle{Phys. Rev. Lett.}} \textbf{115}, 036402
  (2015).

\bibitem{PhysRevX.8.031068}
T Kluge, et~al., Observation of ultrafast solid-density plasma dynamics using
  femtosecond x-ray pulses from a free-electron laser.
\newblock {\em\protect\JournalTitle{Phys. Rev. X}} \textbf{8}, 031068 (2018).

\bibitem{moses_national_2009}
EI Moses, RN Boyd, BA Remington, CJ Keane, R Al-Ayat, The {National} {Ignition}
  {Facility}: {Ushering} in a new age for high energy density science.
\newblock {\em\protect\JournalTitle{Phys.~Plasmas}} \textbf{16}, 041006 (2009).

\bibitem{ellis2020accelerating}
JA Ellis, et~al., Accelerating finite-temperature kohn-sham density functional
  theory with deep neural networks (2020).

\bibitem{Brockherde2017BypassingTK}
F Brockherde, L Li, K Burke, K M{\"u}ller, Bypassing the kohn-sham equations
  with machine learning.
\newblock {\em\protect\JournalTitle{Nature Communications}} \textbf{8} (2017).

\bibitem{Dornheim_PRL_2020_ESA}
T Dornheim, et~al., Effective static approximation: A fast and reliable tool
  for warm-dense matter theory.
\newblock {\em\protect\JournalTitle{Phys. Rev. Lett.}} \textbf{125}, 235001
  (2020).

\bibitem{debroglie}
de~Broglie~L., {\em Nonlinear wave mechanics: A causal interpretation}.
\newblock (Elsevier, Amsterdam), (1960).

\bibitem{Shojai}
A Shojai, F Shojai, About some problems raised by the relativistic form of
  de-broglie–bohm theory of pilot wave.
\newblock {\em\protect\JournalTitle{Physica Scripta}} \textbf{64}, 413 (2006).

\bibitem{GPAW2}
J Enkovaara, et~al., Electronic structure calculations with {GPAW}: a
  real-space implementation of the projector augmented-wave method.
\newblock {\em\protect\JournalTitle{Journal of Physics: Condensed Matter}}
  \textbf{22}, 253202 (2010).

\bibitem{Perdew_LDA}
JP Perdew, A Zunger, Self-interaction correction to density-functional
  approximations for many-electron systems.
\newblock {\em\protect\JournalTitle{Phys. Rev. B}} \textbf{23}, 5048--5079
  (1981).

\bibitem{dornheim_prl16}
T Dornheim, et~al., {Ab Initio Quantum Monte Carlo Simulation of the Warm Dense
  Electron Gas in the Thermodynamic Limit}.
\newblock {\em\protect\JournalTitle{Phys. Rev. Lett.}} \textbf{117}, 156403
  (2016).

\bibitem{PBE}
JP Perdew, K Burke, M Ernzerhof, Generalized gradient approximation made
  simple.
\newblock {\em\protect\JournalTitle{Phys. Rev. Lett.}} \textbf{77}, 3865--3868
  (1996).

\bibitem{PBEsol}
JP Perdew, et~al., Restoring the density-gradient expansion for exchange in
  solids and surfaces.
\newblock {\em\protect\JournalTitle{Phys. Rev. Lett.}} \textbf{100}, 136406
  (2008).

\bibitem{AM05}
R Armiento, AE Mattsson, Functional designed to include surface effects in
  self-consistent density functional theory.
\newblock {\em\protect\JournalTitle{Phys. Rev. B}} \textbf{72}, 085108 (2005).

\end{thebibliography}

\end{document}